How to analyse percentile impact data meaningfully in bibliometrics:

The statistical analysis of distributions, percentile rank classes and top-cited

papers

Lutz Bornmann

Division for Science and Innovation Studies, Administrative Headquarters of the Max Planck

Society, Hofgartenstraße 8, 80539 Munich, Germany; bornmann@gv.mpg.de.

**Abstract**

According to current research in bibliometrics, percentiles (or percentile rank classes) are the most suitable method for normalising the citation counts of individual publications in terms of the subject area, the document type and the publication year. Up to now, bibliometric research has concerned itself primarily with the calculation of percentiles. This study suggests how percentiles can be analysed meaningfully for an evaluation study. Publication sets from four universities are compared with each other to provide sample data. These suggestions take into account on the one hand the distribution of percentiles over the publications in the sets (here: universities) and on the other hand concentrate on the range of publications with the highest citation impact – that is, the range which is usually of most interest in the evaluation of scientific performance.

# 1 Introduction

According to current research in bibliometrics, percentiles (or percentile rank classes) are the most suitable method for normalising the citation counts of individual publications in terms of the subject area, the document type and the publication year (Bornmann, de Moya Anegón, & Leydesdorff, 2012; Bornmann, Mutz, Marx, Schier, & Daniel, 2011; Leydesdorff, Bornmann, Mutz, & Opthof, 2011). Until today, it has been customary in evaluative bibliometrics to use the arithmetic mean value to normalize citation data (Waltman, van Eck, van Leeuwen, Visser, & van Raan, 2011). According to the results from Albarrán, Crespo, Ortuño, and Ruiz-Castillo (2011) (and many other studies), the distribution of citations in every subject category is very skewed, however: "The mean is 20 points above the median, while 9-10% of all articles in the upper tail account for about 44% of all citations" (p. 385). The skewed distribution poses the risk that the citation statistics are dominated by a few highly cited papers (Boyack, 2004; Waltman et al., 2012). This is not possible with statistics based on percentiles. Using percentiles (or percentile rank classes) to normalise citations can therefore give better comparisons of the impact of publications from different subject areas and years of publication and with different document types than normalisation using the arithmetic mean.

The percentile provides information about the impact the publication in question has had compared to other publications (in the same subject area and publication year and with the same document type). Using the distribution of citation counts (sorted in descending order), all the publications from the same subject area and the same year of publication and of the same document type as the publication in question (the reference set) are broken down into 100 percentile ranks. In this study, the maximum value of 100 indicates that a publication has received 0 citations. The lower the percentile rank for a publication, the more citations it has received compared to publications in the same subject area and publication year and with the same document type. The percentile for the respective publication is determined using the distribution of the percentile ranks over all publications. For example, a value of 10 means that the publication in question is among the 10% most cited publications; the other 90% of the publications have achieved less impact. A value of 50 represents the median and thus an average impact compared to the other publications (from the same subject area and publication year and with the same document type).

Up to now, research in bibliometrics has concerned itself primarily with the calculation of percentiles. Some different proposals have been made to calculate percentiles, the intention of the proposals being to find solutions for the problem of small publication sets (Bornmann, in press; Leydesdorff, et al., 2011; Rousseau, 2012; Schreiber, 2012a, 2012b). For example, a reference set may contain a limited number of ten papers. In that case, the highest possible percentile would be based on (9/10) or 90%. Rousseau (2012) proposes defining this highest possible rank as 100% by including the ranked paper in the ranking, and thus considering (10/10) as the highest possible rank (see Leydesdorff, in press). There are only a few isolated attempts to find a meaningful statistical analysis of percentiles in the literature (Bornmann, Mutz, et al., 2011; Bornmann, Schier, Marx, & Daniel, 2011; Boyack, 2004). By categorising percentiles in different percentile rank classes, however, percentiles permit a wide spectrum of analyses of great bibliometric interest.

For example, Leydesdorff and Bornmann (2011) proposed the I3 indicator which is based on percentiles and gives information about the number of publications in a publication set and its citations in a single number (the exact definition is given below). The community picked up on the new indicator I3 shortly after its introduction and began to discuss it (Mutz & Daniel, 2012; Prathap, 2012; Rousseau, 2012; Wagner & Leydesdorff, 2012). Starting from the premise that "complex situations cannot be described by one parameter, one number" (Maurer, 2012, p. 225), this study presents proposals on how percentiles – going beyond I3 – can be analysed meaningfully. Publication sets from four universities are compared with each other to provide sample data. These suggestions take into account on the one hand the distribution of percentiles over the publications in the sets (here: universities) and on the other hand concentrate on the range of publications with the highest citation impact – that is, the range which is usually of most interest in the evaluation of scientific performance.

## 2 Methods

### 2.1 Data base

Publications produced by four universities in German-speaking countries from 2005 to 2009 are used as data. Only the document type "articles" is included in the study. The data was obtained from InCites (Thomson Reuters). InCites (http://incites.thomsonreuters.com/) is a web-based research evaluation tool allowing assessment of the productivity and citation impact of institutions. The metrics (such as the percentile for each individual publication) are generated from a dataset of 22 million Web of Science (WoS, Thomson Reuters) publications from 1981 to 2010. Percentiles are defined by Thomson Reuters as follows: "The percentile in which the paper ranks in its category and database year, based on total citations received by the paper. The higher the number [of] citations, the smaller the percentile number [is]. The maximum percentile value is 100, indicating 0 cites received. Only article types *article, note, and review* are used to determine the percentile distribution, and only those same article types receive a percentile value. If a journal is classified into more than one subject area, the percentile is based on the subject area in which the paper performs best, i.e. the lowest value" (http://incites.isiknowledge.com/common/help/h_glossary.html).

The statistical analyses in this study are based on the one hand on percentiles and on the other on two different percentile rank classes which are often used in bibliometrics. (i) With PR(6) the percentiles are categorised in six rank classes (Bornmann & Mutz, 2011). This approach is already in use as the evaluation scheme in the Science and Engineering Indicators of the US National Science Foundation (National Science Board, 2012). In this scheme, the focus is on publications cited more frequently than the median. The six percentile rank classes are aggregated as follows:

(1) <50% (papers with a percentile greater than the 50$^{th}$ percentile),

(2) 50% (papers within the [50$^{th}$; 25$^{th}$[ percentile interval),

(3) 25% (papers within the [25$^{th}$; 10$^{th}$[ percentile interval),

(4) 10% (papers within the [10$^{th}$; 5$^{th}$[ percentile interval),

(5) 5% (papers within the [5$^{th}$; 1$^{st}$[ percentile interval),

(6) 1% (papers with a percentile equal to or smaller than the 1$^{st}$ percentile).

(ii) With PR(2) the percentiles are categorised in two rank classes. As it has now established itself as standard on an institutional level to designate those publications in a set which are among the 10% most cited publications in their subject area (publication year, document type) as highly cited publications (Bornmann, de Moya Anegón, et al., 2012; Tijssen & van Leeuwen, 2006; Tijssen, Visser, & van Leeuwen, 2002; Waltman, et al., 2012), the following two rank classes are formed:

(1) <90% (papers with a percentile greater than the 10$^{th}$ percentile),

(2) 10% (papers with a percentile equal to or smaller than the 10$^{th}$ percentile).

## 2.2 Statistical significance tests

Bibliometric data used for a university in an evaluation study is usually a random sample from a population. The population is formed by all the publications from one university and their citations, which are recorded in the relevant literature databases. Using statistical significance tests, it is possible to verify whether it is highly probable that the results for the random sample are valid for the population of publications (Bornmann, Mutz, Neuhaus, & Daniel, 2008). Statistical significance tests either assume the random selection of a sample from the population or it is necessary to select a suitable strategy for selecting the random sample which excludes systematic errors and dependencies with regard to the question raised in the study (here: the differences in bibliometric indicators between universities). A random selection was not made for this study. Instead, publication periods and citation windows were chosen from which we can assume that there would be no systematic advantages or disadvantages for a university. If a test which looks at the difference between two universities with regard to the citation impact of their publications turns out to be statistically significant, it can be assumed that the difference has not arisen by chance, but can be interpreted beyond the data at hand.

As in this study relatively high publication figures are available for the individual universities and the results of statistical significance tests are also dependent on the number of cases in a study, an increased significance level of $p<.001$ is used. Due to the high numbers of publications, the power of the statistical tests might be high, so that statistically non-significant parameters really reflect zero effects. Where possible in the following, standard errors and corresponding confidence intervals are calculated which tell us how much estimated parameters vary from sample to sample.

As there is a statistically significant difference between the percentiles and normal distribution (tested with the skewness and kurtosis test for normality), non-parametric statistics are used for their analysis.

## 3    Results

Table 1 shows, broken down by publication year, the number of publications for the four universities which are included in this study. Of a total 17,537 publications most come from Univ 1 (n=5,921) and Univ 2 (n=5,598). Univ 3 has the fewest publications (n=2,074). Table 2 shows some summary statistics for the percentiles per university. The percentiles are distributed from a value close to 0 (minimum; here the most cited publications) to a value of 100 (maximum; here non-cited publications). The standard deviation is similar for all the universities with a value of around 30. While on average Univ 1 (m=47.46), Univ 2 (m=47.92) and Univ 4 (m=48) show a citation performance which is equivalent to the mean in their subject, (that is close to 50), Univ 3's performance (m=37.87) is significantly better. The lower the percentile median for a university, the higher the median citation impact of the publications.

Figure 1 shows the distribution of percentiles for the 4 universities in a graph. The violin plot (Hintze & Nelson, 1998) shows a marker for the median of the percentiles (50% of the values lie above or below the value), a box indicating the interquartile range (the first quartile – 25% of the values – and the third quartile – 75% of the values), and spikes extending to the upper and lower adjacent values. Furthermore, plots of the estimated kernel density are overlaid. As the diagram shows, Univ 3 not only has a better median citations performance (i.e. a lower median percentile) than the other universities; the university also stands out with a lower number of less cited publications (percentile values close to 100) and an interquartile range shifted towards lower percentile values (that is, towards higher numbers of cited publications).

**3.1  Results based on percentiles**

Figure 2 uses box plots to show the distribution of percentiles for the 4 universities by publication year. The dotted line through the graph indicates an average citation impact of 50. The median citation impact over all the publication years is given in the label for each university. As the medians (like the means in Table 2) show for each university, the values for Univ 1 (med=42.5), Univ 2 (43.67) and Univ 4 (med=41.87) indicate a median citation impact for the publications, which is roughly equivalent to the average in one subject area (publication year and document type). Only the value for Univ 3, med = 29.75, is significantly above the mean (in terms of citation impact).

The box plots in Figure 2 consist of a box, the outer limits of which characterise the interquartile range (see above). The cross in the centre of the box designates the median for one university in one year. The position of the median in the box conveys an impression of the skewness of the values. A comparison of publication years shows that particularly the percentiles for 2009 have been shifted towards higher values: a series of publications from this year have either not been cited or very little. As this pattern is visible primarily for the most recent publication year, the citing window before the cites were recorded is apparently too short (from the publication in 2009 to the end of 2010), to allow reliable statements to be made about the impact of the publications (see also the results by Radicchi & Castellano, 2011). This result indicates the necessity to take account of publication years which are already several years in the past in evaluation studies.

Figure 2 (in the labels for the universities) shows the result of pairwise comparisons of the impact of publications by the 4 universities. These individual comparisons were carried out following the H-test by Kruskal and Wallis (Sheskin, 2007, pp. 981-1006), which indicated a statistically significant difference between the medians of the percentiles for all the universities ($\chi^2_3 = 174.65$, $p < .01$). The individual comparisons provide information about the differences between two universities to which this general finding (the statistically significant result from all the universities) can be attributed. The repetition of the statistical test on each pairwise comparison makes it necessary to apply a Bonferroni correction to the significance level. As the result of the individual comparisons in the figure shows, there is a statistically significant difference between Univ 3 and the other universities. We can therefore consider with great probability that it holds true for all the publications that the median impact of the publications from Univ 3 is above the median impact of publications from the other universities.

## 3.2 Results based on PR(6)

The first stage of the analysis was to calculate I3 values for the 4 universities based on PR(6). I3 can formally be written as

$$I3 = \sum_i x_i * f(x_i)$$

in which $x_i$ denotes the percentile rank class $i$, and $f$ the number of publications with this value (Leydesdorff & Bornmann, 2011; Wagner & Leydesdorff, 2012). With I3, citation impact in a publication set is weighted in accordance with the percentile rank class of each publication. The indicator combines output and citation impact in one figure similar to the h index (Hirsch, 2005) (h number of publications with at least h citations). However, the advantage of I3 over the h index is that it uses citation impact measures normalized with percentiles, which makes I3 values comparable over subject areas. As the indicator is dependent on the publication output (this dependence was also intended in the definition) and we have two universities in this study, Univ 3 and Univ 4 which have significantly fewer publications than Univ 1 and Univ 2, a percentage value is calculated for each university in addition to I3. The percentage value shows which share the I3 value for a university has reached if the theoretically maximum possible value of I3 (all the publications from one university are in the class 1%) is used as a benchmark. Measured using I3, Univ 1 (I3=12,519) and Univ 2 (I3=11,745) performed better than Univ 4 (I3=8,378) and much better than Univ 3 (I3=5,169). However, the I3 values correlate clearly with the publication numbers. Measured on the theoretical I3, which is the maximum a university can reach with the number of publications, Univ 3 (42%) has a (significantly) better value than Univ 1, Univ 2 and Univ 4 (each with 35%).

The second stage in the analysis used PR(6) in this section for a differentiated view on the individual percentile impact classes of the publications. Figure 3 shows the differences in the various classes between the four universities based on PR(6). It shows the percentage share of publications from each university in the individual classes. For example, 43.52% of the publications from Univ 1 belong in class <50% and 1.42% in class 1% in their subject area. For Univ 2, 44.64% of publications are in class <50% and 1.16% in class 1%. The categorisation of the publications in six classes and the graphic representation of the percentage values (Cox, 2004) in the figure permits differences in the performance of the universities to be looked at more closely. To which publications (e.g. publications with very high or very low citations) can differences over all the publications be primarily attributed? In which classes does the number of publications from one university deviate from what one would expect from a random selection of publications in a database?

First, the $\chi^2$-test was used to determine how statistically significant the difference between the universities is in terms of their citation impact (Sheskin, 2007). The results showed a statistically significant variation ($\chi^2_{15} = 177.17$, $p < .01$) which agrees with the results in section 3.1. Using the $\chi^2$-decomposition it is now possible to test the contribution of each cell (here: bar) to the overall $\chi^2$-value of 177.17. This contribution is given for each cell (over the bar) in Figure 3. The higher the value is, the more the number of publications in the cell (the observed value) deviates from the expected value (the expected value for a cell is calculated from the line and column totals in each case) (Sheskin, 2007). To judge by the height of the $\chi^2$-values in the figure, the statistically significant variation between the universities is attributable primarily to the little cited publications (class <50%; with an $\chi^2$-total for the whole line of 72.9) and Univ 3 (with an $\chi^2$-total for the total column of 146.6). Compared to the other universities, Univ 3 with 31% has considerably fewer publications in the class <50% than the other universities with percentage values significantly over 40% ($\chi^2$-value=63.2). Furthermore, for Univ 3 there are considerably more publications in the class 5% (9.31%) and the class 1% (2.89%) than for the other universities ($\chi^2$-values=28.8 and 24.7). The advantages in the performance by Univ 3 are therefore in the small proportion of little cited and the high proportion of much cited publications.

Using the percentages values in Figure 3 it is not only possible to compare the universities with each other, but also to compare the percentage values of each university with expected values which would result from a random selection of publications from a database (Bornmann, 2010; Bornmann, Mutz, et al., 2011; Zitt & Lahatte, 2009). For example, we can expect for each university that 1% of their publications belong in the class 1% and 4% in the class 5%. Thus, for example, with 6% (Univ 1), 6.16% (Univ 2), 9.31% (Univ 3) and 5.55% (Univ 4) all four universities published (significantly) more class 5% papers than the expected value of 4%. For the class 50%, for all four universities there are percentage values which roughly agree with the expected value of 25% (between 23.3% and 25.02%). The expected values form unambiguous reference values for the percentage values in the PR(6) which can be interpreted as a difference (deviation) from the reference value.

### 3.3 Results based on PR(2)

Finally, the percentiles were categorised in two rank classes in this study. The publications in the top 10% in their subject area (publication year, document type) (class 10%), and the remaining publications (class <90%) are each collected into one class. As described above, the publications in the class 10% can be seen as the meaningful range of publications in a set (measured by citation impact). Figure 4 shows the number of publications from the 4 universities which are among the 10% most cited publications in their subject area. Error bars are shown for each bar showing a 95% confidence interval. The solid line shows the expected value of 10% und the dashed line the percentage of the class 10% for all the universities. The percentage values given in the label for each university are the proportion of class 10% publications for this university.

As the percentage figures in the labels of Figure 4 show, the proportion for each university is more than 10%. As there is no confidence interval cutting across the solid line, we can with great certainty assume that all the universities have a higher proportion in the class 10% than might be expected on the basis of a random sample from the population of publications in a database. Particularly for Univ 3 with a share of 21%, there are significantly more publications in the class 10% than would be expected from a random selection.

Using PR(2) as an example, a regression model was calculated and the results are presented in the following section. It could also have been calculated for percentiles and PR(6) (but then with a different model specification depending on the data type). The benefit of the regression model is that as well as the group variable (here the universities) and the citation impact (here PR(2)) other independent variables can be included, whose effect on the result can be controlled. In recent years, a number of factors have been discussed which can have an effect on the citation impact of individual publications (as well as the subject area, publication year, and document type which have already been taken into account in the percentiles) (Bornmann & Daniel, 2008; Bornmann, Schier, Marx, & Daniel, 2012). As examples of these influencing factors, the number of pages in a publication and the number of authors have been used in this study (see Table 3 (i)). It is assumed that long publications and publications with a large number of authors are more likely to be cited than short publications and publications with just a few authors. As Table 3 (i) shows, the data set for example includes publications with pages averaging around 10 and a maximum number of pages of 172. Furthermore, the regression model can take account of dependencies in the data set which arise because university staff have published jointly. Jointly published publications are present more than once in the data set.

To identify citation impact differences between the four universities, we used a multiple logistic regression model (Hosmer & Lemeshow, 2000; Mitchell, 2012). This model is appropriate for the analysis of dichotomous (or binary) responses. Dichotomous responses arise when the outcome is the presence or absence of an event (Rabe-Hesketh & Everitt, 2004). In this study, the binary response is coded 0 for a publication in the class <90% and 1 for a publication in the class 10%. In the model, the four universities, the number of pages, and the number of authors are the independent variables, and the dichotomous responses (PR(2)) are the dependent variable. The violation of the assumption of independent observations caused by including the same publication more than once for different universities is taken account of in the logistic regression model by using the "cluster" option in Stata (Hosmer & Lemeshow, 2000, section 8.3; StataCorp., 2011). As only those publications can be included in the model for which there are no missing values for any of the variables, 14,647 publications can be used of the total of 17,537.

The results of the regression model are shown in Table 3 (ii). There is a statistically significant influence for all the variables. Given the factors assumed to be exerting an influence on the citation impact, these results show that with a higher number of pages and a larger group of authors, citation impact is expected to be higher. Table 3 (iii) shows the results for the 4 universities of pairwise comparisons of marginal linear predictions which were calculated subsequent to the regression model. Pairwise comparisons using Bonferroni's adjustment for the p-values and confidence intervals are calculated to test which universities' citation impact exhibits a statistically significant difference (Mitchell, 2012). Unlike the pairwise comparisons which were dealt with in section 3.1, these are results which are adjusted by the factors influencing citation impact described above. Consistent with the results previously given in Table 3 (iii), only Univ 3 is statistically significantly different from all the other universities; none of the other pairwise comparisons is statistically significant.

Figure 5 shows the differences between the universities with a graph as the result of the regression model. It shows the adjusted predictions with 95% confidence intervals of the number of publications from the 4 universities, which are in the 10% most cited publications in their subject area. The predictions are adjusted for the length of the publications, the number of authors and the dependence in the data set, which has arisen from the assignment of the same publication to different universities.

# 4    Discussion

While a number of publications have looked at the calculation of percentiles and the assignment of publications to percentile rank classes (see the introduction), only a few have considered the analysis of this data, leading to meaningful results in evaluation studies. This study discusses several approaches to the analysis of percentiles, PR(6) and PR(2) (publications which belong to the 10% most cited publications in their subject area, document type and publication year). Significance tests were presented which are able to indicate statistically meaningful results (such as the $\chi^2$-test and the $\chi^2$-contribution) and results visualised in graphics which show the distribution of data meaningfully (such as the violin plots). Furthermore, single-number indicators, such as I3 and the number of publications in class 10% were presented. Percentiles and percentile rank classes offer – as this study shows – an abundance of options for data analysis with which research institutions (such as universities) can be compared not only with each other, but with which a unit can also be compared with expected values.

Sample data from 4 universities for which there are very large data sets available has been used for this study. The analyses which are proposed here can however also be applied to other sets, such as publications by individual scientists, research groups and from certain journals. However, it should be ensured that there are sufficient publications for analysis. In order to present the results of an evaluation study in this way, as here based on PR(6), for example, it is necessary for the individual percentile rank classes to contain sufficient publications. A sufficient number of publications should not be a problem for research groups; however, at individual scientist level it is to be expected that there are only sufficiently large numbers available for scientists who have been active for many years.

Generally speaking, a bibliometric evaluation study should always make use of several indicators in order to give a differentiated view of the scientific performance of a unit. To this end, different approaches are presented in this study. As the h index has acquired such significance in the evaluation of research over recent years, the author would like to draw attention to the indicator "proportion of class 10% publications" in one set as a better alternative to the h index. Similarly to the h index (Bornmann & Daniel, 2007; Hirsch, 2005), the proportion of class 10% publications provides information about productivity and citation impact in a single figure: It gives the number of publications, which – measured on their citation impact – could achieve greater significance.

A benefit of the "proportion of class 10% publications" compared to the h index is that an arbitrary threshold is not used to select the significant publications in a set. Numerous publications about the h index have already criticised the use of an arbitrary threshold (see the overview in Waltman & van Eck, 2012). "For instance, the h-index could equally well have been defined as follows: A scientist has an h-index of h if h of his publications each have at least 2h citations and his remaining publications each have fewer than 2(h+1) citations. Or the following definition could have been proposed: A scientist has an h-index of h if h of his publications each have at least h/2 citations and his remaining publications each have fewer than (h+1)/2 citations" (Waltman & van Eck, 2012, p. 408). In contrast to the h index, no arbitrary threshold is used to calculate the proportion of class 10% publications, but a standard bibliometric method is used to identify significant publications in a set (see above). As well as using the standard procedure the proportion of class 10% publications offers the following two crucial advantages over the h index: (i) As percentiles are normalised for subject area and time, proportions of class 10% publications can be compared between any subject matters or periods. This is not possible with the h index. (ii) As it can be expected with a publication set that 10% of the publications are in the 10% of the most cited publications (in their subject area), the actual proportion of these publications can be compared with the expected value of 10% (Bornmann, de Moya Anegón, et al., 2012). It is also possible to verify whether the performance is above or below expectations. That is not possible with the h index either.

Some proposals for an analysis of percentile data that can lead to meaningful results in evaluation studies have been made in this study. It would be very desirable if the approaches suggested here were tested and developed with other data sets and other research units.

Table 1. Number of publications in the years 2005 to 2009 for four universities

| Publication year | Univ 1 | Univ 2 | Univ 3 | Univ 4 | Total |
|---|---|---|---|---|---|
| 2005 | 1,311 | 925 | 277 | 743 | 3,256 |
| 2006 | 1,282 | 988 | 364 | 765 | 3,399 |
| 2007 | 1,274 | 1,167 | 449 | 793 | 3,683 |
| 2008 | 1,029 | 1,238 | 470 | 813 | 3,550 |
| 2009 | 1,025 | 1,280 | 514 | 830 | 3,649 |
| Total | 5,921 | 5,598 | 2,074 | 3,944 | 17,537 |

Table 2. Summary percentile statistics for four universities

| University | Number of publications | Minimum (most cited papers) | Maximum (papers with 0 citations) | Mean | Standard deviation |
|---|---|---|---|---|---|
| Univ 1 | 5921 | 0.02 | 100 | 47.46 | 32.39 |
| Univ 2 | 5598 | 0.01 | 100 | 47.92 | 32.45 |
| Univ 3 | 2074 | 0.01 | 100 | 37.87 | 30.59 |
| Univ 4 | 3944 | 0.04 | 100 | 48 | 33.55 |
| Total | 17537 | 0.01 | 100 | 46.60 | 32.62 |

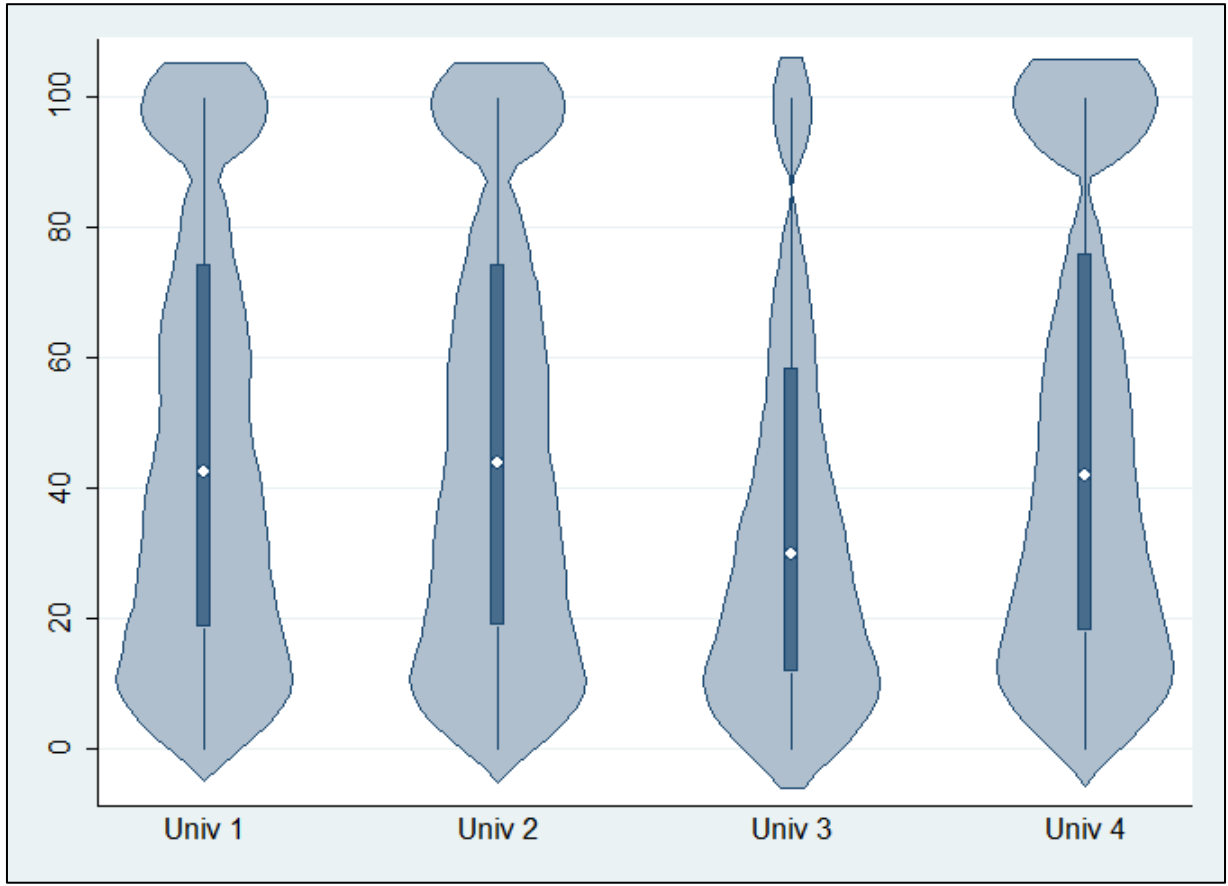

Figure 1. Percentile distributions for four universities. The violin plots show a marker for the median of the percentiles, a box indicating the interquartile range, and spikes extending to the upper- and lower-adjacent values. Furthermore, plots of the estimated kernel density are overlaid.

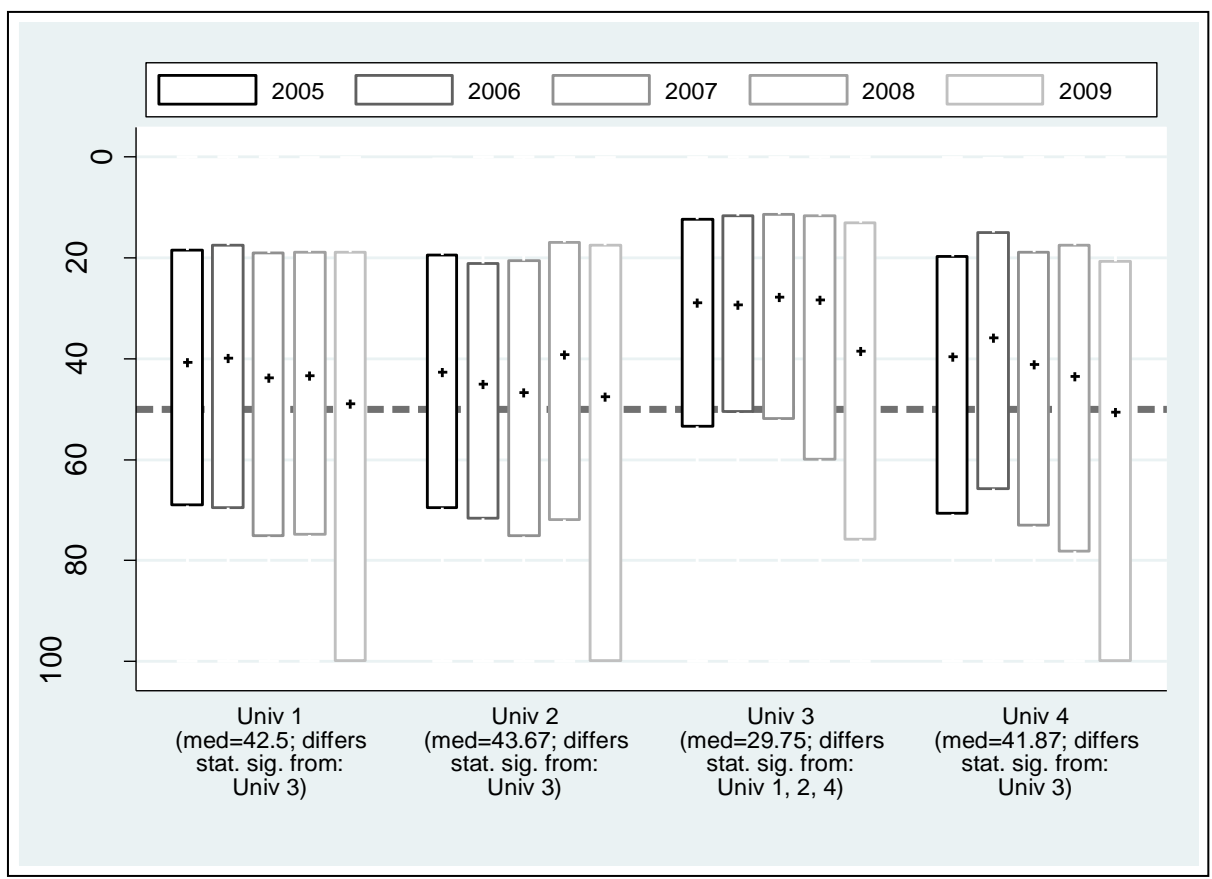

Figure 2. Percentile distributions for four universities broken down by publication year. The solid line in the figure indicates an average citation impact of 50. The cross in the box designates the median for one university in one year. The outer margins of the box indicate the first quartile (25% of the values and the third quartile (75% of the values). The median of the percentiles for each university across all years is given in the labels; furthermore, those universities are named here from which a university differs statistically significantly in its citation performance.

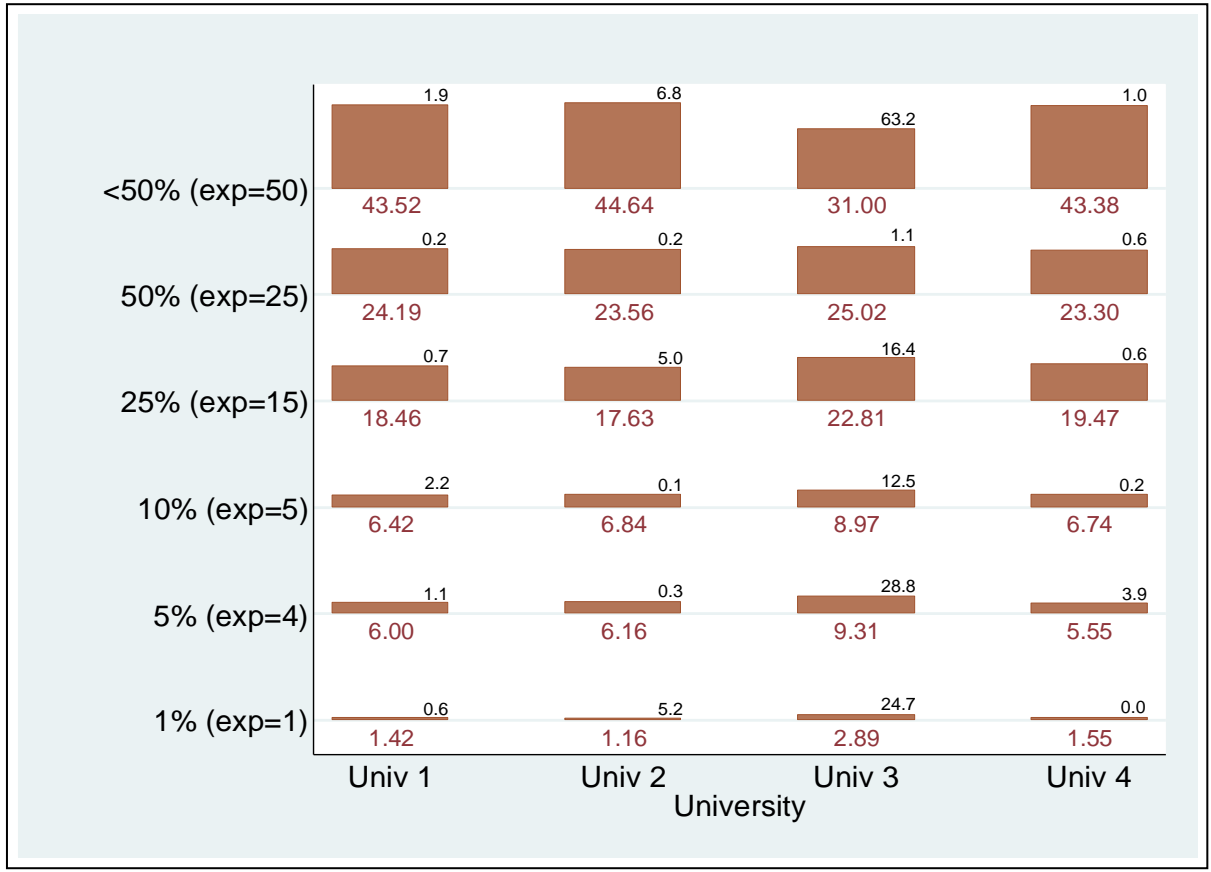

Figure 3. Differences between the universities measured by PR(6). The figure shows the percentage of a university's publications within the different classes using the bar height and as indicated below the bar. The $\chi^2$-contribution of each bar (cell) to the statistically significant overall $\chi^2$-value ($\chi^2_{15} = 177.17$, $p < .01$) is located above the bar. The labels of PR(6) show the percentage (exp) in brackets, which can be expected for a bar (cell) if the publications are randomly selected.

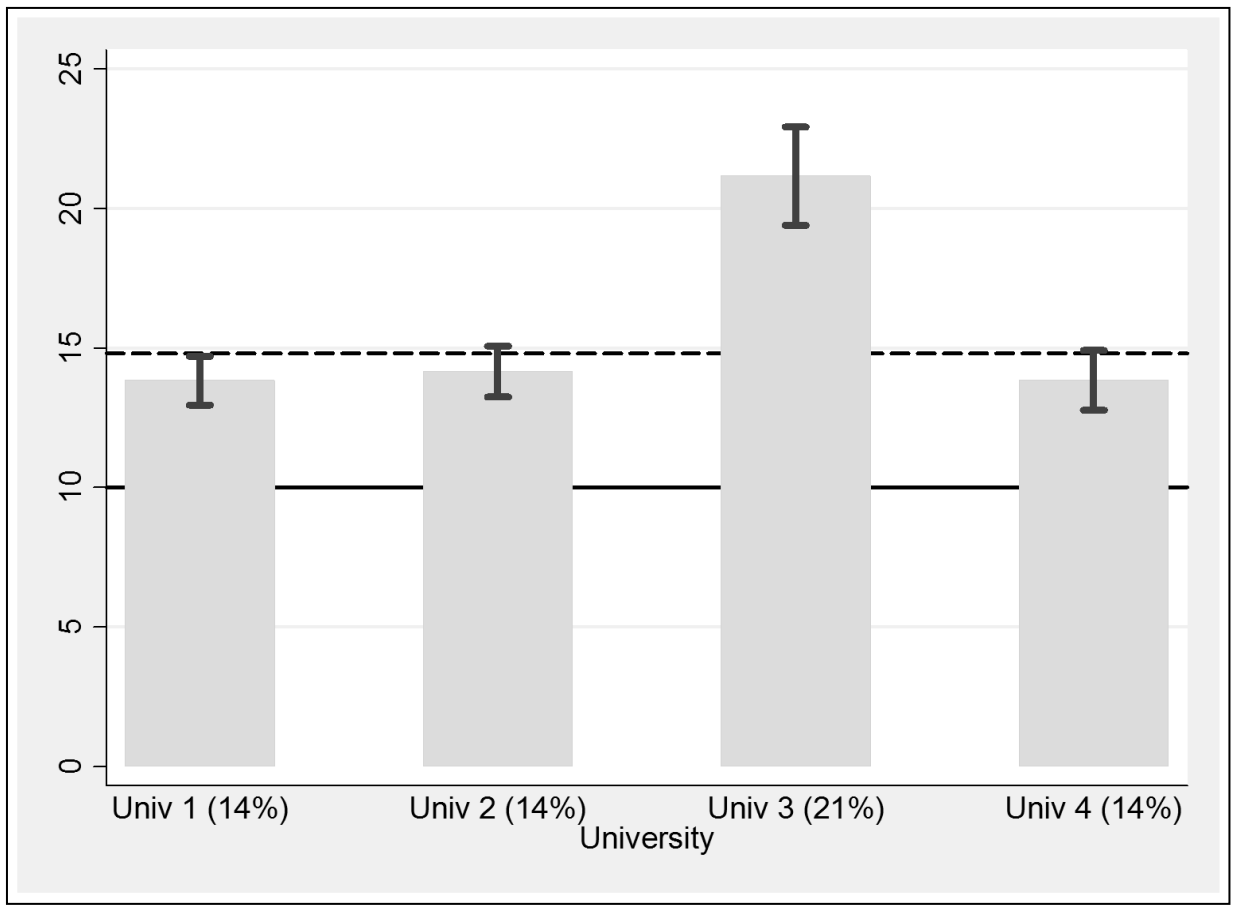

Figure 4. Number of publications from the 4 universities which are in the 10% of the most cited publications in their subject area. The error bars are given for each bar showing a 95% confidence interval. The solid line shows the expected value of 10% und the dashed line the percentage of the 10% most cited publications for all the universities. The percentage values given in the label for each university are the proportion of 10% most cited publications for this university.

Table 3. Binary regression model calculating differences between the four universities taking into account the length of publications, the number of authors and the dependency in the data set which arises from the assignment of the same publication to different universities.

(i) Characteristic values for the variables used in the binary regression model (n=14647)

| Variable | Mean | Standard deviation | Minimum | Maximum |
|---|---|---|---|---|
| PR(2) | 0.14 | 0.34 | 0 | 1 |
| University | | | | |
|   Univ 2 | 0.33 | 0.47 | 0 | 1 |
|   Univ 3 | 0.12 | 0.33 | 0 | 1 |
|   Univ 4 | 0.21 | 0.41 | 0 | 1 |
| Number of pages | 10.15 | 6.71 | 1 | 172 |
| Number of authors | 5.28 | 3.24 | 1 | 22 |

Note: Standard deviations are not adjusted for clustering

(ii) Results of the binary regression model

| Variable | Coefficient | t statistics |
|---|---|---|
| University | | |
|   Univ 2 | 0.14* | 2.31 |
|   Univ 3 | 0.45*** | 6.09 |
|   Univ 4 | 0.14 | 1.97 |
| Number of pages | 0.02*** | 5.73 |
| Number of authors | 0.14*** | 19.76 |
| Constant | -3.00*** | -39.16 |
| Number of publications | 14647 | |
| Number of unique publications | 14461 | |

Notes: * $p < 0.05$, *** $p < 0.001$

(iii) Pairwise comparisons of marginal linear predictions

| Pair | Contrast | t statistics |
|---|---|---|
| Univ 2 vs Univ 1 | 0.14 | 2.31 |
| Univ 3 vs Univ 1 | 0.45*** | 6.09 |
| Univ 4 vs Univ 1 | 0.14 | 1.97 |
| Univ 3 vs Univ 2 | 0.31*** | 4.16 |
| Univ 4 vs Univ 2 | 0.00 | 1.00 |
| Univ 4 vs Univ 3 | -0.31** | -3.67 |

Notes: ** $p < 0.01$, *** $p < 0.001$

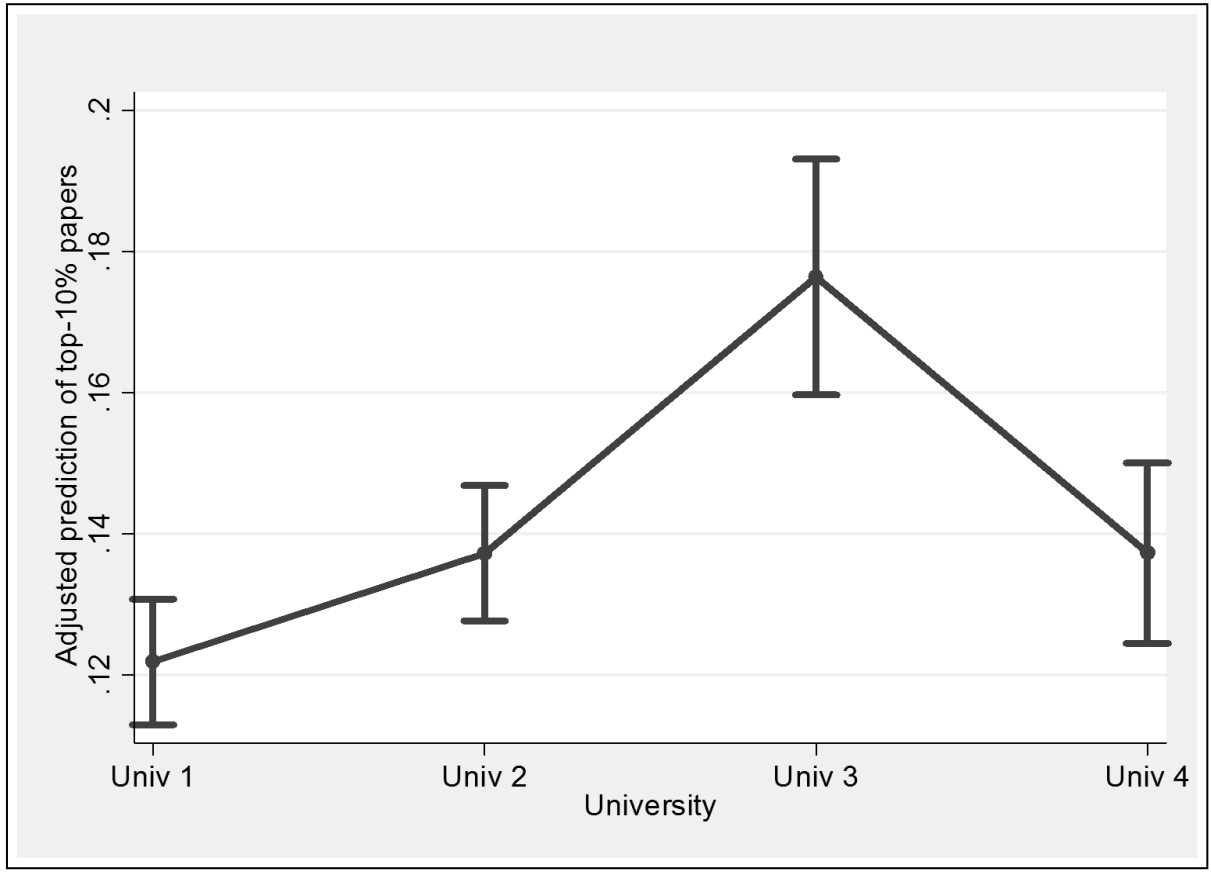

Figure 5. Adjusted predictions with 95% confidence intervals of the number of publications from the 4 universities, which are in the 10% most cited publications in their subject area. The predictions are adjusted for the length of the publications, the number of authors and the dependence in the data set, which has arisen from the assignment of the same publication to different universities.